\documentclass[prb,showpacs,floatfix,twocolumn]{revtex4}
\usepackage{graphicx}
\usepackage{amssymb}
\usepackage{dcolumn}
\usepackage{bm}
\begin{document}

\title{Avalanche size distributions in mean field plastic yielding models}
\author{E. A. Jagla}
\affiliation{Centro At\'omico Bariloche and Instituto Balseiro (UNCu)\\ 
Comisi\'on Nacional de Energ\'{\i}a At\'omica, (8400) Bariloche, Argentina}
\date{\today}

\begin{abstract}

I discuss the size distribution ${\cal N}(S)$ of avalanches occurring at the yielding transition of mean field (i.e., Hebraud-Lequeux) models  of amorphous solids. 
The size distribution follows a power law dependence of the form: ${\cal N}(S)\sim S^{-\tau}$. However
(contrary to what is found in its depinning counterpart) the value of $\tau$ depends on details of the dynamic protocol used. For random triggering of avalanches I recover the $\tau=3/2$ exponent typical of mean field models, which in particular is valid for the depinning case. However, for the physically relevant case of external loading through a quasistatic increase of applied strain, a smaller exponent (close to 1) is obtained. This result is rationalized by mapping the problem to an effective random walk in the presence of a moving absorbing boundary.

\end{abstract}

\maketitle

\section{Introduction}

The yielding transition (at zero temperature) of an amorphous solid material occurs when the externally applied shear stress $\sigma$
overpasses a critical value $\sigma_c$. For $\sigma < \sigma_c$ the system is blocked and remains in this state indefinitely. As  $\sigma > \sigma_c$ the system enters a plastic flow regime in which the strain in the system increases linearly with time. This is referred to as a plastic flow regime. The transition at $\sigma = \sigma_c$ is the yielding transition. At $\sigma_c$ the system is in a critical state, and the dynamics proceeds via avalanches of all sizes, characterized by its size distribution ${\cal N}(S)$.

The previous scenario has been confirmed by a variety of experimental measurements,\cite{hohler,roberts,cloitre,mobius,becu} as well as analytical and numerical techniques.\cite{maloney1,lemaitre,karmakar,salerno,maloney2,gimbert,talamali,baret,martens,picard,lin,lin2,ferrero} From all these sources,  the idea has emerged that plasticity proceeds through the destabilization and rearrangement of discrete units at given spatial positions (sometimes referred to as shear transformation zones).\cite{argon,falk}
Rearrangement of a unit at a given spatial position has an effect (through direct elastic interaction) on the stability of other units at different locations.
It is thus clear that the form of the elastic interactions between different regions of the material play a central role in the problem. In this respect, it is crucial to realize that the elastic interaction kernel has alternating signs in different spatial regions.\cite{picard4}
The destabilization of a given site can thus have both a destabilizing or stabilizing effect on some other site. 

The previous description shares many features with the well known case of depinning transitions of elastic manifolds on disordered energy landscapes.\cite{fisher,kardar} However, a crucial difference steps from the fact that for depinning, destabilization of a given site always produces a destabilizing effect on any other site, namely the interaction kernel is positive definite.
The mean field description of the depinning transition is well studied, and is known to produce a critical size distribution of avalanches of the form ${\cal N}(S)\sim S^{-3/2}$. It has to be mentioned that this value of $\tau$ stands for (at least) two quite different protocols used to trigger avalanches in the system. In one case ({\em random loading}) one site is chosen at random and the local stress is increased until the site destabilizes. In the second case ({\em uniform loading}) the stress on the system is increased uniformly over all sites, until the first site becomes destabilized. In both cases $\tau=3/2$ is found for depinning.

I will show that in the case of mean field plastic yielding models, the value of $\tau$ that is obtained is dependent on the loading mechanism applied. While for random destabilization the value $\tau=3/2$ is obtained, 
uniform loading produces a smaller value of $\tau$, close to 1. The reason of this difference is the much larger stress that is added to the system in the uniform loading protocol compared to the random loading case. I present the model to be studied in the next Section. An analytical treatment of the problem is given in Section III, and results of numerical simulations are presented in Section IV.  Section V contains the conclusions.

\section{The model}

Hebraud and Lequeux\cite{hebraud} considered a mean field model of the plastic yielding transition. 
The model considers a real valued variable $y_i$, representing the stress at each spatial position of the system. \cite{nota1}
The coupling among $y_i$ variables is mean field like, i.e., any change in $y_i$ affects equally all other $y$'s. 
I discuss here a small variation of the original HL model, in the limit of quasistatic loading. 

I consider $N$ stresses $y_i$, $i=1,...N$, which define the instantaneous configuration of the system. I will refer to the $N$ variables $i$ alternatively as "sites", or "particles".
The configuration is stable as long as all $y_i$ are lower than a plastic threshold value, taken for simplicity to be uniform, of value 1. It will be more convenient to describe the model in terms of the variables $x_i\equiv 1-y_i$. Each $x_i$  represents the amount to additional stress necessary to destabilize site $i$. The dynamics proceeds according to the following rules.

(a) On a stable configuration, an instability is produced (either by random or uniform loading), and an avalanche starts. Suppose $i$ is the site that becomes unstable. 


(b) $x_i$ is increased a quantity $(1+k)u$, where $u$ is a random quantity, of order 1 (below I take $u$ to be exponentially distributed: $p(u)=\overline u ^{-1} \exp(-u/\overline u)$, $u>0$), and $k$ is a non-conservation parameter (in such a way that criticality is expected as $k\to 0$).

(c) All $x_j$ are decreased a quantity $\eta/\sqrt N-u/N$, where
$\eta$ is a Gaussian variable, with zero mean and fixed standard deviation $\eta_0$.

(d): Steps (b-c) are repeated over all destabilized sites, \cite{nota2} until all sites become stable (this defines the end of the avalanche). Then I return to (a).

The instantaneous stress $\sigma$ in the model is defined as $\sigma=\frac 1N\sum_i (1-x_i)$. The size $S$ of an avalanche is conveniently defined as the sum of all $u$ values generated during the avalanche.
Note that according to the previous rules, an avalanche of size $S$ generates a reduction of stress in the system of value $kS/N$. As $k\to 0$ the average size of avalanches goes to infinity as $1/k$. This corresponds to a broad distribution of avalanches sizes in this limit, which corresponds to criticality. For finite $k$, there is typically a maximum size $S_{max}$  of the avalanches that are generated.

The previous rules indicate that each site reaching the $x=0$ border is reinserted at $x=(1+k)u$, then $x=0$ is an absorbing border. Each unstable site also produces random kicks (of intensity $\eta/\sqrt N$) plus a systematic decrease (of value $-u/N$) on all other sites. In terms of a fictitious time that counts the number of sites that reach the $x=0$ border, these two terms can be described as diffusive, and convective, respectively.

In the thermodynamics (large $N$) limit, the equilibrium distribution of $x$ values is characterized by a function $P(x)$, such that $NP(x)\Delta x$ gives the number of sites with $x$ within the small interval $\Delta x$. 
The form of $P(x)$ is determined as a balance between the sites that reach $x=0$ and are thus redistributed, and the change in $x$ that each site reaching $x=0$ produces. In an equilibrium situation the total change $\Delta P$ produced by a single particle being destabilized and reinserted must vanish, and this means that $P(x)$ satisfies the equation (for $k\to 0$)

\begin{equation}
\Delta P=0=\frac DN \frac{\partial^2 P}{\partial x^2}+\frac 1 N\frac{\partial P}{\partial x}+\frac 1N{(DP'(0)+P(0))}p\left (x\right )
\label{eq}
\end{equation}
with $D=\eta_0^2/2$ of step (c) above, and $P(x)=0$ for $x<0$.
The three terms of the r.h.s. represent respectively the effect of diffusive and convective variations of $x$ (the two terms in the change of $x_j$ at step (c)), and the effect of reinserted sites (step (b)). The amplitude of the reinserted term, namely $DP'(0)+P(0)$ takes into account that sites can reach the $x=0$ border due to the diffusive, or convective evolution.

There is an important difference in the equilibrium form of $P(x)$ when $D=0$, or $D>0$. In the first case (in which the previous model describes depinning) $P(x)$ is immediately obtained from (\ref{eq}) as

\begin{equation}
P(x)={P(0)}\int_x^\infty p (x)dx
\end{equation}
In particular, $P(x)$
goes to a non-zero value when $x\to 0$. For $D>0$, due to the diffusive term and the absorbing boundary condition, $P(x)$ goes to zero at $x=0$. Moreover, the particle rate across the $x=0$ border due to the diffusive term is proportional to $dP/dx|_{x=0}$, and for this quantity to be finite, the form of $P(x)$ must be $P(x)\sim Ax$ for small $x$. This is a mean field results that does not hold for short range interactions, where a form $P(x)\sim x^\theta $ with a non-trivial $\theta$ is expected.

For $k\to 0$, $P(x)$ is independent of the form of the destabilizing mechanism (random or uniform loading) since at criticality the stress redistribution caused by direct external loading is negligible compared with reaccommodations during avalanches (this is not true for finite $k$, see the Appendix).
The form of $P(x)$ for different values of $D$ are presented in Section IV.

\section{Results for avalanche statistics}

In order to calculate the avalanche statistics, the continuous description provided by the $P(x)$ function is not sufficient. In fact, what this distribution tells is that each single site reaching $x=0$, produces changes in all other $x$ such that {\em on average} an additional site reaches $x=0$, i.e., a conservative critical situation is reached. However, in order to describe avalanche distribution we must consider fluctuations. 

\subsection{Random loading}

It is convenient at this point to describe the dynamics of the system in term of a discrete "time unit" corresponding to one step of points (b) and (c) of the dynamical protocol described above. Namely, at each discrete time, one kick is given to every particle according to (c), and one destabilized particle is reinserted according to (b).
Let us call $n_i$ the cumulative number of destabilized particles until time $t\equiv i$. The condition for an avalanche started at $t=0$ to survive until time $t=i$ is that $n_i\ge i$. The avalanche stops at the moment in which $n_i< i$. 

Since the sites carrying different values of $x$, once destabilized, are reinserted through a random process (implied by the random values of the $u$ parameter) the exact values of $x_i$ are locally uncorrelated. 
Due to this uncorrelated behavior, the stochastic process $n_i$ corresponds to a Poisson process with a unitary rate per time step. Defining the stochastic process $m_i$ as $m_i=n_i-i$, we see that the avalanche lasts until the first time $i_0$ in which $m_{i_0}=0$.
Note that as one single site is reinserted at each time step, $i_0$ measures also the avalanche size, i.e., $S\equiv i_0$.

In the large $i$ limit $m_i$ is a symmetric random walk, with probability distribution $p(m_i)$ given by

\begin{equation}
p(m_i)=\frac 1{\sqrt {2\pi i }}\exp{\left ( -\frac{m_i^2}{2i}   \right )}
\label{pdem}
\end{equation}
and the previous description corresponds to its survival probability in the presence of an absorbing boundary at the origin.
This surviving probability goes as $t^{-1/2}$. The size distribution of avalanches is given by the probability density of the random walk being absorbed at time $t$, which gives ${\cal N}(S)\sim t^{-3/2}$, i.e., $\tau=3/2$. I emphasize that the previous method shows this result is valid for random loading, both in the case $\eta_0=0$ (depinning), and $\eta_0 > 0$ (plastic yielding) as it only depends on the fact that the number of sites reaching the instability threshold is an uncorrelated stochastic process with constant rate.

\subsection{Uniform loading}

In the uniform loading case, in order to start an avalanche we look for the smallest $x$ value in the system, say $x_{min}$, and add this additional load to the system. This is equivalent to say that all $x$'s in the system are reduced in the quantity $x_{min}$. The question is if this shifting of the $P(x)$ distribution may bring some observable effect. We will see that the answer depends on whether we are considering the case
$\eta_0=0$ or $\eta_0>0$. 

For $\eta_0=0$ (depinning case) the form of $P(x)$ tends to a constant as $x\to 0$, i.e, $P(x)\sim P(0)$ for small $x$. 
When shifting the distribution by $x_{min}$ to start the avalanche, we modify $P(x)$ to a new $\widetilde P(x)$
given by $\widetilde P(x)=P(x+x_{min})\sim P(0)$ (to leading order). This means that the form of $P(x)$ is not greatly affected by the $x_{min}$ shift. In fact, results of numerical simulations described below show that both loading protocols produce the same avalanche distribution ${\cal N}(S)\sim S^{-3/2}$ in the $\eta_0=0$ case.

However, the situation is different in the case $\eta_0>0$. Now $P(x)\sim  Ax$ for small $x$, and a shift of the distribution by $x_{min}$ transforms it to $\widetilde P(x)=P(x+x_{min})\sim  Ax+ Ax_{min}$, namely the probability distribution gets a constant correction near $x=0$ which qualitatively modifies its form. This produces an important change in the avalanche size distribution as I will show now.

Suppose we have the distribution $P(x)= Ax+A_0$. The linear part
generates a flux of particles through the $x=0$ border characterized by the Poissonian stochastic process $n_i$. The average rate of this process is easily seen to be equal to $AD$, and since we have one particle arriving per unit time, we conclude that $A=1/D$. The constant term $A_0$ produces the arrival at $x=0$ of some additional particles, that contribute to the process. Let $d_i$ be the number of additional particles arrived at $x=0$ at or before time $i$, originated in the additional constant term $A_0$. Now the stochastic process to be considered in order to determine the size of the avalanches is $\widetilde m_i=n_i-i+d_i$, i.e, the avalanche survives as long as $\widetilde m_i\ge 0$. We need to characterize the contribution $d_i$ from $A_0$.
Due again to the fact that sites that contribute to the constant $A_0$  are totally uncorrelated, $d_i$ is a (cumulative) Poissonian process, however now its average rate depends on time. The time dependence of the average rate can be calculated simply noticing that it gives the average number of particles absorbed at $x=0$, starting from a distribution with constant value $A_0$ for all $x>0$. The average number $\overline d_i$
can be obtained by direct integration of the diffusion equation, and the result is
\begin{equation}
\overline d_i = 2A_0\sqrt {\frac{Di}{\pi}}
\label{cuatro}
\end{equation}
The condition for an avalanche to survive until time $i$ is then
$\widetilde m_i=m_i+2A_0\sqrt {{Di }/{\pi }}\ge 0$. In other words, the original condition $m_i\ge 0$ is now
replaced by $m_i\ge -2A_0\sqrt {Di/\pi}$, i.e., the absorbing boundary that was originally at $i=0$ now
retracts in time, as $\sim i^{1/2}$. 

\begin{figure}[h]
\includegraphics[width=8cm,clip=true]{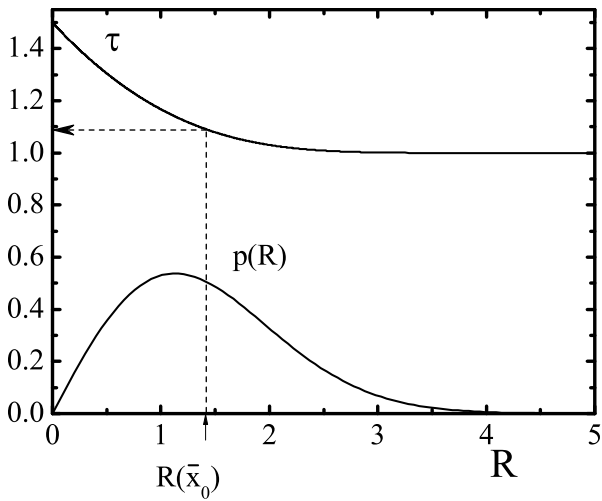}
\caption{The $\tau(R)$ function obtained from the mapping to the survival probability of a random walk of variance $\sigma^2=t$, in the presence of an absorbing wall that recedes as -$R\sqrt t$. I also show the distribution of $R$ values according to its probability distribution $p(R)$, and the value of $\tau\simeq 1.09$ obtained for its average value.
}
\label{f0}
\end{figure}

The effect of such a moving absorbing boundary condition on a random walk has been studied in the literature (see for instance Refs. \cite{sato,redner,bray}). The important point that makes the problem analytically solvable is that the position of the absorbing boundary has the same dependence in time that the internal diffusive dynamics, 
and this allows to transform the problem to a single ordinary differential equation.\cite{nota3} The survival probability is thus observed to remain a power law in time, but with a non-universal exponent. The value of $\tau$ that depends on the coefficient 
$R\equiv 2A_0\sqrt {D/\pi}$ of the square root time behavior in Eq. \ref{cuatro} (assuming a normalized underlying random walk, as in Eq. \ref{pdem}).
It is  found that $\tau=3/2$ for $R=0$, and $\tau\to 1$ when $R\to \infty$. We see the full form of this dependence in Fig. \ref{f0}.

\begin{figure}[h]
\includegraphics[width=8cm,clip=true]{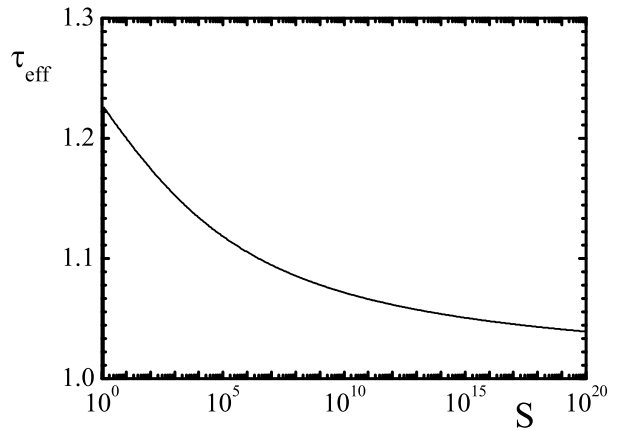}
\caption{The local exponent $\tau_{\tiny\mbox {eff}}$ as a function of the avalanche size $S$. The value of $\tau_{\tiny\mbox {eff}}$ is a very weakly decreasing function of $S$.
}
\label{suma_alfa}
\end{figure}

In order to calculate the actual value of $R$, we must determine what value of $A_0$ is appropriate in our case. It turns out that we do not have a unique value of $A_0$, but a continuous distribution. In fact, given a starting distribution $P(x) =Ax$, the lowest $x$ value, which defines $x_{min}$ is itself a random variable, with distribution

\begin{equation}
p(x_{min})={Ax_{min}}\exp{\left (-\frac{Ax^2_{min}}{2}\right )}
\label{pdexmin}
\end{equation}
We can write $x_{min}=x_0/\sqrt A$, where $x_0$ follows the normalized distribution
\begin{equation}
p(x_0)={x_0}\exp{\left (-\frac{x^2_0}{2}\right )}
\end{equation}
The values of $A_0$ are obtained from $A_0=Ax_{min}=\sqrt A x_0$, giving for $R$ the value (remember that $A=1/D$):

\begin{equation}
R = {\frac{2 x_0}{\sqrt \pi}}
\end{equation}
We can estimate a typical value of $\tau$ as the one corresponding to $x_0$
equal to its average value ($\overline x_0= \sqrt{\pi/2}$). This gives $R=\sqrt{2}$, that entered in Fig. \ref{f0} gives $\tau\simeq 1.09$.
However we must notice that
different avalanches are produced with different $x_0$ values, and the final distribution of avalanche sizes has to be calculated as a composition from avalanches taken from distributions with different $\tau$ values
according to

\begin{equation}
{\cal N}_{TOT}(S)=\int_{S_{min}}^{\infty} d\tau p(\tau)N(S(\tau))
\label{ntot}
\end{equation}
where $p(\tau)=p(x_0)dx_0/d\tau$ (taken from data in Fig. \ref{f0}) is the probability of different values of $\tau$, and ${\cal N}(S(\tau))=C_{\tau}S^{-\tau}$, with $C_{\tau}$ a normalizing constant. Taking the minimum size of avalanches to be $S_{min}\equiv 1$, it results $C_{\tau}=(\tau-1)$, and a numerical evaluation of the expression (\ref{ntot}) produces a non-perfect power law, that can be 
characterized by an effective, local exponent $\tau_{\tiny\mbox {eff}}$ defined as
\begin{equation}
\tau_{\tiny\mbox  {eff}}\equiv\frac {d\log [N_{TOT}(S)]}{d\log S}
\end{equation}
The result of this evaluation can be seen in Fig. \ref{suma_alfa}.
As we move to larger values of $S$, $\tau_{\tiny\mbox {eff}}$ is more dominated by the terms with the smallest $\tau$ in Eq. (\ref{ntot}), and $\tau_{\tiny\mbox {eff}}$ decreases. It has to be noticed however, that for this effect to be appreciable we must really move to extremely large values of $S$. As a rule of thumb, we can say that we expect a $\tau$ between 1.1 and 1.2 to be observed.

\section{Numerical Simulations}

I will present here results of numerical simulation of the model described in Section II, that will confirm the scenario described in the previous Section. In all the simulations presented the values of $u$ are taken from an exponential distribution with mean value of 1, namely $p(u)=\exp{(-u)}$.
In this case, in the thermodynamic limit the equilibrium form of $P(x)$ can be readily found by integrating Eq. \ref{eq}, and is given (for $k\to 0$) by

\begin{equation}
P(x)=\frac{1}{1-D}\left ( e^{-x}-e^{-x/D}     \right )
\label{eq_sol}
\end{equation}

\begin{figure}[h]
\includegraphics[width=8cm,clip=true]{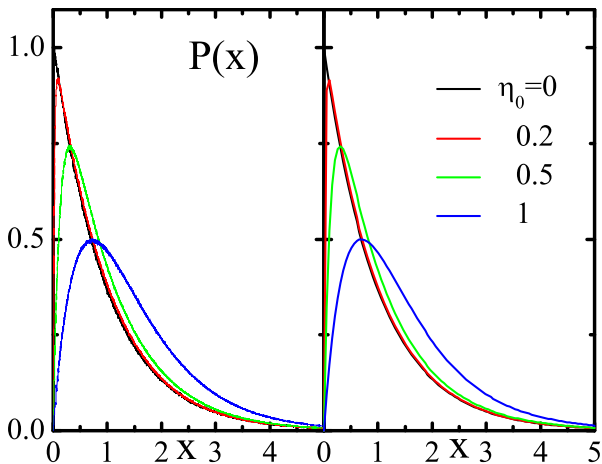}
\caption{Form of the function $P(x)$ for different values of $\eta_0$, in the $k\to 0$ limit. 
The left panel contains results of simulations with $N=10000$. The right panel shows the analytical limit for $N\to \infty$.
The crucial characteristic of these curves is the behavior near $x=0$. Note that $P(x)\sim x^0$ for $\eta_0=0$, whereas $P(x)\sim x^1$
for $\eta_0\ne 0$.
}
\label{f1}
\end{figure}

The form of $P(x)$ for different values of $\eta_0$ at criticality ($k\to 0$) is shown in Fig. \ref{f1}. Note that it must be $D<1$ for the solution to exist.
As it was explained before, the main feature of these curves is their behavior near $x=0$. We see that that $P(x)\sim x^0$ for $\eta_0=0$, whereas $P(x)\sim x^1$ for $\eta_0\ne 0$.

\begin{figure}[h]
\includegraphics[width=8cm,clip=true]{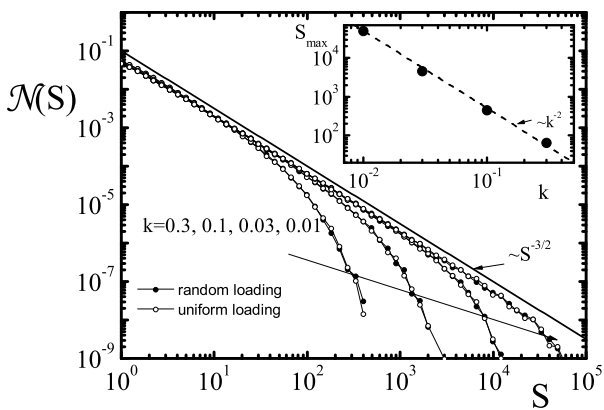}
\caption{Simulations in a model with $N=10^5$, and $\eta_0=0$ (i.e., depinning case). Avalanche size distribution for decreasing values of $k$. Results for the two different loading protocols (uniform and random) are shown, and they are seen to be equivalent. The distribution observed has the form ${\cal N}(S)\sim S^{-3/2}\exp (-S/S_{max})$. The dependence of $S_{max}$ on $k$ is shown in the inset, and is $S_{max}\sim k^{-2}$.
}
\label{f2}
\end{figure}

Next, I present results for the avalanche size distributions.
Results in Fig. \ref{f2} correspond to the depinning model ($\eta_0=0$). They were obtained in systems with $N=10^5$. It is seen that the size distribution ${\cal N}(S)$ is a power law ${\cal N}(S)\sim S^{-3/2}$ that is cut-off at large avalanche size at some characteristic typical maximum size $S_{max}$, that increases as $k$ decreases. We can obtain the increase law of $S_{max}$ with $k$ as follows. An avalanche of size $S$ generates a reduction in the systems stress of value $Sk$. This quantity is (on average) compensated by the stress increase in the loading stage. For the $\eta_0=0$ case, this increase is given on average by $\overline u$, which is equal to one in our case. So we get
$\overline S= k^{-1}$. Also, $\overline S$ is related to $S_{max}$ through $\overline S \sim S_{max}^{2-\tau} $, which gives $S_{max}\sim k^{-\frac 1{2-\tau}}$, and in the present case with $\tau=3/2$, we get 
$S_{max}\sim k^{-2}$, that is well satisfied by the results in Fig. \ref{f2}.
Note that the results for $\eta_0=0$ show no difference for the two different driving protocols, namely random or uniform loading, according to the arguments given in the previous Section.

\begin{figure}[h]
\includegraphics[width=8cm,clip=true]{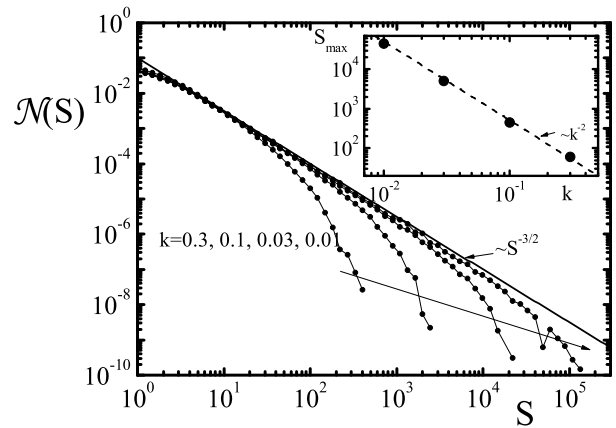}
\caption{Same as previous figure but for $\eta_0=1$ in the random loading case. The results are seen to be similar to the case of depinning ($\eta_0=0$), with an avalanche size distribution with an exponent $\tau=3/2$, and a large size cut off $S_{max}\sim k^{-2}$.
}
\label{f3}
\end{figure}

Now we move to the analysis of the plastic yielding case ($\eta_0\ne 0$).
In this case, the two loading protocols generate different results. For random loading we find a power law distribution with $\tau=3/2$ (Fig. \ref{f3}), equivalent to that found in the depinning case. 

\begin{figure}[h]
\includegraphics[width=8cm,clip=true]{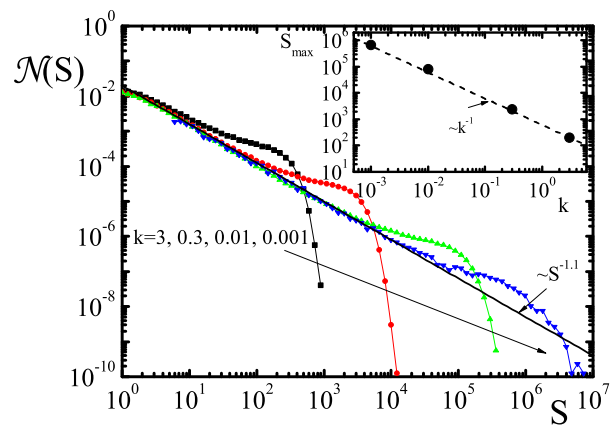}
\caption{Simulations for $\eta_0=1$, $N=10^5$ in the uniform loading case. It is seen that as $k$ decreases, a power law with a value smaller than the standard $\tau=3/2$ value is observed. It is also apparent the existence of a rather strong overshoot at the range on the largest avalanches observed.
}
\label{f4}
\end{figure}

In the case of uniform loading the results are different, as expected from the analysis of the previous section. In Fig. \ref{f4} we see the decay of number of avalanches with size. 
On the range of sizes displayed in Fig. \ref{f4}, a decaying exponent $\tau\simeq 1.1$ can be estimated as $k$ is decreased. 
I notice the rather strong overshoot of the size distributions near $S_{max}$, that I analyze in the Appendix.

\begin{figure}[h]
\includegraphics[width=8cm,clip=true]{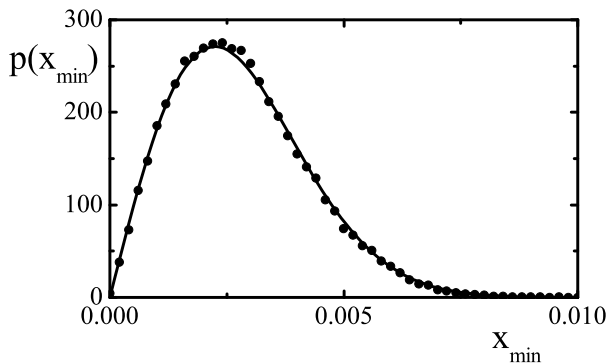}
\caption{(a) The distribution of values of $x_{min}$ in a simulation with $N=10^5$ and $\eta_0=1$, $k=0.1$. The continuous line is the analytical expectation given by Eq. \ref{pdexmin}.
}
\label{f5}
\end{figure}

To gain some additional insight on the results in Fig. \ref{f4}, I present some additional analysis concerning the avalanche size and the values of $x_{min}$ at which they are triggered. In Fig. \ref{f5} we see the distribution of values of $x_{min}$ obtained along a simulation with $N=10^5$, $k=0.1$. We see that 
the distribution follows very closely the distribution given by Eq. (\ref{pdexmin}), which in turns originates in the linear distribution of $x$ values near $x=0$. 

\begin{figure}[h]
\includegraphics[width=8cm,clip=true]{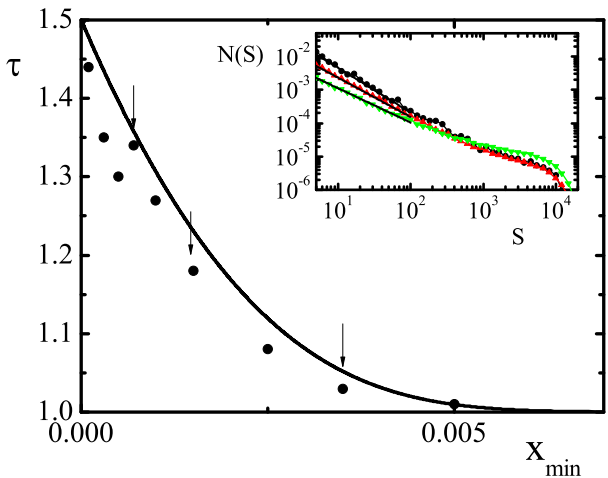}
\caption{The $\tau$ exponent of partial distribution of avalanches in the simulation of the previous figure, classified according to the value of $x_{min}$ at which avalanches are triggered. 
In the inset we see an example of the distributions and the low size ranges (thick straight lines) used to extract the $\tau$ exponent. The three curves in the inset correspond to the three points in the main plot indicated by arrows.
}
\label{f6}
\end{figure}

In addition, if avalanches are independent, and they follow the results from the theory of the previous section, we should expect a different decaying exponent for avalanches if they are classified according to the value of $x_{min}$ at which they were triggered. 
In order to check for this, in the inset to Fig. \ref{f6} I show partial size distributions of avalanches restricted to given intervals  of values of $x_{min}$ that trigger them. If we concentrate in the low size part of this plots, the results follow a trend compatible with the theoretical analysis: 
avalanches started with smaller $x_{min}$ have a distribution with a decaying exponent close to 3/2, while those with larger $x_{min}$ produce a distribution with a slower decaying exponent.
There seems to be however systematically lower values in the simulations compared with the theoretical results. 

There is an additional effect in the uniform loading plastic yielding case that deserves to be mentioned. For depinning, or random loading plastic yielding, I have argued that the mean size of the avalanches must satisfy the relation $\overline S k\sim 1$, since the average external load between avalanches is order 1. In the case of uniform loading in plastic yielding, the average loading between avalanches is order $Nx_{min}\sim \sqrt N$. So in this case we have $\overline S k\sim \sqrt N$, or $\overline S \sim \sqrt N/k$. Introducing the size of the largest avalanches in the system $S_{max}$ we get $S_{max} \sim  (\sqrt N/k)^{1/(2-\tau)}$. This indicates that the size of the largest avalanches in the system are not completely determined by the value of $k$. There is also a direct dependence on $N$. In particular, at any fixed value of $k$, $S_{max}$ becomes arbitrarily large as $N\to \infty$.

\section{Conclusions}

In this paper I have studied the avalanche size distributions in mean field models of the plastic depinning transition. The main conclusion is that contrary to the case of elastic depinning where the mean field exponent is well defined and given by $\tau=3/2$, here the result depends on details of the driving mechanism. I have shown that for plastic depinning, the $\tau=3/2$ exponent is recovered in the case of random loading, where each avalanche is started by making unstable a random site. However, the more standard loading mechanism corresponding to increase uniformly the stress in all the system until an instability is obtained, produces a significantly lower exponent $\tau\simeq 1.1-1.2$. The origin of this effect was elucidated by mapping the present problem to the survival probability of a random walk in the presence of a moving absorbing boundary. In addition, this mapping provides insight into the size crossover for cases in which the system is not right at criticality.

The present results call the attention on the precise destabilization mechanism used in different models of plastic depinning. It is likely that the effects discussed here remain in more realistic cases (not mean field) where a more realistic symmetry and interaction range of the elastic kernel is considered. The kind of variation of the $\tau$ exponent in those cases remains to be elucidated.

\section{Acknowledgements}

I thank Alberto Rosso for stimulating discussions, and P. Le Doussal for bringing to my attention Refs. \cite{sato,redner}. This research was financially supported by Consejo Nacional de Investigaciones Cient\'{\i}ficas y T\'ecnicas (CONICET), Argentina. Partial support from grant PICT-2012-3032 (ANPCyT, Argentina) is also acknowledged.

\appendix

\section{The approach to criticality, and the large size overshoot}

It has already been noticed (in connection with data in Fig. \ref{f4}) the important size effect observed at large avalanche sizes, where an excess number of avalanches (compared with the results for random loading, Fig. \ref{f3}) is observed. 
In order to study this effect in more detail, it is not enough to consider the critical ($k=0$) case, but it is necessary to analyze the form in which the system evolves when a finite $k$ is reduced towards 0.

I restrict here to the plastic depinning case ($\eta_0\ne 0$). The modifications in the balance equation (\ref{eq}) produced by a finite $k$ are different in the cases of uniform, or random loading. 
In the case of random loading, a finite $k$ produces (according to the rules of Section II) that the reinsertion does not occur at the position $u$, but at $u(1+k)$. This alters the argument of $p$ and the normalization of the last term in (\ref{eq}). In addition, for finite $k$, the average size of the avalanche $\overline =1/k$ is finite, and we must take into account that 1 every $1/k$ particles is destabilized by the random mechanism that initiates the avalanche. As this random selection is made on the actual distribution $P(x)$, this produces a negative term $-kP/N$ per destabilized particle in the balance equation, that finally reads

\begin{equation}
0=\frac DN \frac{\partial^2 P}{\partial x^2}+\frac 1N\frac{\partial P}{\partial x}-\frac {kP}N+\frac{DP'(0)}{N(1+k)}p\left (x/(1+k)\right )
\end{equation}

In the case of uniform loading, we still have the change $u\to u(1+k)$ in the reinsertion term, but now
the uniform loading mechanism produces a shift of $P$ in a quantity $x_{min}$ every $\overline S$ destabilized particles. This gives a correction $\delta P$ in the balance equation that is equal to $\delta P=(dP/dx)dx=(dP/dx)(x_{min}/\overline S)=(dP/dx)(k/N)$ where the last equality follows (on average) from the balance between the stress increase during loading ($x_{min}$), and the stress decrease during avalanche ($k\overline S/N$). Finally, the balance equation for uniform loading takes the form

\begin{equation}
0=\frac DN \frac{\partial^2 P}{\partial x^2}+\frac{1+k}N\frac{\partial P}{\partial x}+\frac{DP'(0)}{N(1+k)}p\left (x/(1+k)\right )
\label{eq_unif}
\end{equation}

For our choice $p(x)=\exp(-x)$ the solutions to both equations are

\begin{equation}
P(x)=\frac{e^{-\frac{x}{1+k}}-e^{-\frac{x(1+\sqrt{1+4kD})}{2D}} }        {1+k-\frac{2D}{1+\sqrt{1+4kD}}}
\label{eq1}
\end{equation}
for random loading, and 

\begin{equation}
P(x)=\frac{e^{-\frac{x}{1+k}}-e^{-\frac{x(1+k)}{D}} }        {1+k-\frac D{1+k}}
\label{eq2}
\end{equation}
for uniform loading.

In the limit $N\to\infty$, the distribution of avalanches is controlled by the form of these expressions near $x=0$. Close to criticality, we must search for the leading terms in an expansion in powers of $k$. For random loading, linearizing Eq.  (\ref{eq1}) near $x=0$, and writing the coefficient to first order in $k$ we obtain

\begin{equation}
P(x)=\frac{x}D \left(1-k(1-D)\right)
\end{equation}
where we see that as $k\to 0$, $P(x)$ tends to the critical form $P(x)=x/D$. The finite value of $k$ provides a cut off in the size distribution of avalanches. In fact, in the mapping to the random walk problem, the linear in $k$ term of the previous expression corresponds to an absorbing moving wall at $m_i=ki/D$ (in the notation of Section III). 
This problem can be analytically solved in the large avalanche size limit, and provides an avalanche distribution given by 
\begin{equation}
{\cal N}(S)\sim S^{-3/2} \exp (-Sk^2/D^2)
\end{equation}
This gives a size cutoff $S_{max}$ that increases a $1/k^2$ as $k\to 0$, and results that fit nicely the curves in Fig. \ref{f3}.

For uniform loading instead, it is easily verified that Eq. (\ref{eq2}) is independent of $k$ to first order in $x$.\cite{nota4} This means that in order to obtain an avalanche cutoff we must expand to second order in $x$. The result (to first order in $k$ in the coefficient of $x^2$) is

\begin{equation}
P(x)= \frac{x}{D}-\frac 1D \left(D+1+k (1-D)\right)x^2
\label{px}
\end{equation}

\begin{figure}[h]
\includegraphics[width=8cm,clip=true]{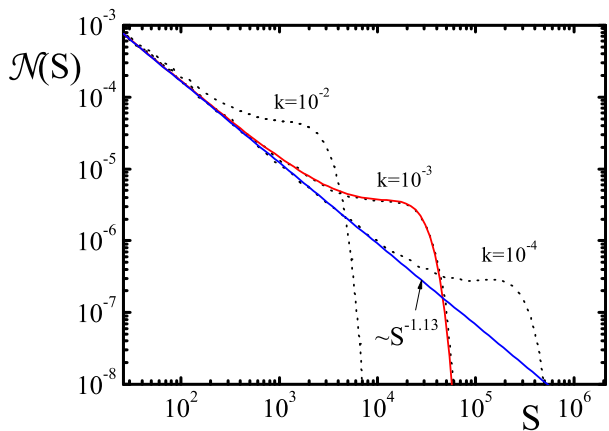}
\caption{Dotted lines: Size distribution of avalanches described by the absorbing of a random walk $m(t)$ by a wall located at $b(t)=-x_{min}t^{1/2}+kt^{3/2}$. Results are presented for different $k$ values. The values of $x_{min}$ are taken from their expected distribution in our problem (Eq. (\ref{pdexmin})). The continuous line is a fitting to the analytical expression (\ref{magic}), using a single average $x_{min}$ value.
}
\label{fapp}
\end{figure}

This form of $P(x)$ allows to explain in detail the behavior observed in Fig. \ref{f4} for the uniform loading case. In fact, when $k=0$ we are at criticality, and we can describe the avalanche distribution through the model of Section III, of the survival of a random walk, where the loading to $x_{min}$ corresponds to the presence of a retracting absorbing wall at $\sim -x_{min}\sqrt t$. The linear in $k$ term in Eq. (\ref{px}) corresponds to an additional current of particles through the absorbing wall that has to be subtracted (because it has a negative coefficient) from the total. The number of absorbed particles from this term until time $t$ is given by $-4k \frac{(1-D)}{\sqrt{D\pi}}t^{3/2}$. It represents an additional "forward movement" of the absorbing wall that is ultimately responsible for the decaying of the random walk in a finite time (i.e, avalanches have a maximum value). 
We end up with the problem of an unbiased random walk $m(t)$ with an absorbing boundary $b(t)$ moving with the law 
\begin{equation}
b(t)=-x_{min}t^{1/2}+kt^{3/2}
\label{bi}
\end{equation}
(some non-essential coefficients have been set to 1 in this expression).
It is straightforward to numerically simulate this process, finding the size of the avalanches as the first time $t_0$ at which $m(t_0)=b(t_0)$, namely when the walk is absorbed, and then making statistics on the values of $t_0$. The results are shown in Fig. \ref{fapp}. Simulations are presented for different $k$ values, whereas $x_{min}$ are taken from its known distribution (Eq. (\ref{pdexmin})). The results give a nice verification of the form of the cut off that was observed in the full simulations of Fig. \ref{f4}.

I notice that an analytical solution to the survival probability of a random walk in the presence of the absorbing wall as in Eq. (\ref{bi}) is not known. However, the form of the cutoff can be heuristically determined by inserting Eq. (\ref{bi}) in the asymptotic form expected for a solution of the diffusion equation of the random walk at large $x$, namely $\sim \exp (-x^2/2t)$, which gives for the cut off the form

\begin{equation}
\sim \exp{\left ( -k^2S^2/2+x_{min}kS    \right )}
\label{magic}
\end{equation}
We can thus expect the form of ${\cal N}(S)$ to be given by
\begin{equation}
{\cal N}(S)\sim S ^{-\tau}\exp{\left ( -k^2S^2/2+x_{min}kS    \right )}
\label{magic}
\end{equation}
where $\tau$ is given as a function of  $x_{min}$ in Section II. In Fig. \ref{fapp} we see that this gives a remarkably good fitting of the numerical results. From Eq. (\ref{magic}) it is clear that the intensity of the overshoot becomes larger as $x_{min}$ increases. \cite{nota7}
Note also that the scaling of $S_{max}$ as $1/k$ is immediate form expression (\ref{magic}).


\begin{references}
\bibitem{nota1} Note that stresses and strains in a full description would require the consideration of rank-2 tensors, instead. However, the scalar description seems to capture most of the phenomenology of the yielding transition.
\bibitem{nota2} The present dynamics assumes that once destabilized, sites are committed to be reinserted according to step (b), i.e., they cannot be re-stabilized by an ulterior kick.
This corresponds to the so-called  "Model $B$" in the description of Ref. \cite{lin}. The results presented below will not be valid for the "Model $A$" in Ref. \cite{lin}.
\bibitem{nota3} This differential equation can be written in a form that corresponds to finding the ground state energy $\epsilon$ of a quantum particle in a parabolic potential $\sim x^2$ limited by a hard wall located at $x=-A_0$. It is found that the value of $\tau$ is given by $\tau=3/4+\epsilon/2\hbar\omega$, and goes from $\tau=3/2$ to $\tau=1$ when $A_0$ moves from $A_0=0$ to $A_0\to\infty$.
\bibitem{nota4} Although this is shown explicitly here for the particular form of $p(x)$ used, it must be stressed that is a general property of Eq. \ref{eq_unif}, for arbitrary forms of $p(x)$.
\bibitem{nota7} Eq. (\ref{magic}) seems to indicate that for $x_{min}=0$ the cut off is purely of the form $\sim \exp(-k^2S^2)$. However numerical simulations show that even in this case, a linear in $S$ term in the exponent is necessary to fit the results. This indicates that Eq. (\ref{magic}) is certainly not a rigorous formula.
\end{references}
\end{document}